\documentclass[journal=jacsat,manuscript=article]{achemso}
\usepackage{color}
\usepackage{graphicx}
\usepackage{amsmath}
\usepackage{xcolor}
\usepackage{comment}
\usepackage{hyperref}
\usepackage{booktabs}
\usepackage{amssymb}
\usepackage[normalem]{ulem}

\usepackage{subfigure}
\usepackage{soul}
\title[Escaping Flatland]
  {Investigating 3D Atomic Environments for Enhanced QSAR}
\author{William McCorkindale}
\affiliation{Cavendish Laboratory, University of Cambridge, Cambridge CB3 0HE, United Kingdom}

 \author{Carl Poelking}
 \affiliation{Department of Chemistry, University of Cambridge CB2 1EW, United Kingdom}

\author{Alpha A. Lee}
\email{aal44@cam.ac.uk}
\affiliation{Cavendish Laboratory, University of Cambridge, Cambridge CB3 0HE, United Kingdom}
\alsoaffiliation{PostEra Inc., 2 Embarcadero Center, San Francisco, CA 94111, USA}

\begin{document}

\begin{abstract}
Predicting bioactivity and physical properties of molecules is a longstanding challenge in drug design. Most approaches use molecular descriptors based on a 2D representation of molecules as a graph of atoms and bonds, abstracting away the molecular shape. A difficulty in accounting for 3D shape is in designing molecular descriptors can precisely capture molecular shape while remaining invariant to rotations/translations. We describe a novel alignment-free 3D QSAR method using Smooth Overlap of Atomic Positions (SOAP), a well-established formalism developed for interpolating potential energy surfaces. We show that this approach rigorously describes local 3D atomic environments to compare molecular shapes in a principled manner. This method performs competitively with traditional fingerprint-based approaches as well as state-of-the-art graph neural networks on pIC$_{50}$ ligand-binding prediction in both random and scaffold split scenarios. We illustrate the utility of SOAP descriptors by showing that its inclusion in ensembling diverse representations statistically improves performance, demonstrating that incorporating 3D atomic environments could lead to enhanced QSAR for cheminformatics.

%Methods that do account for 3D shape typically require a heuristic to align molecules and consider the similarity between the shapes of entire molecules. We describe a novel alignment-free 3D QSAR method using Smooth Overlap of Atomic Positions (SOAP), a well-established formalism developed for interpolating potential energy surfaces. This approach rigorously describes local 3D atomic environments to compare molecular shapes in a principled manner. This method performs competitively with traditional fingerprint-based approaches as well as state-of-the-art graph neural networks on pIC$_{50}$ ligand-binding prediction in both random and scaffold split scenarios. We demonstrate that its inclusion in ensembling diverse representations is statistically superior to ensembling others, and show that incorporating 3D atomic environments could lead to enhanced QSAR for cheminformatics.
\end{abstract}

\maketitle

\section{Introduction}
Predicting physical properties or bioactivity from molecular structure –- quantitative structure–activity relationships (QSAR) modelling -- underpins a large class of problems in drug discovery. Being able to computationally evaluate  molecular properties, from solubility to protein-ligand binding affinity, is vital for medicinal chemists to rationally design and prioritise drug candidates for synthesis and testing. While statistical and machine learning (ML) approaches have been developed for QSAR since the 1970s, significant amount of innovation has occurred in the space of models -- from classical ML methods such as random forest and support vector machines, to the latest technologies based on deep neural networks. Central to any machine learning methodology is the way molecules are described within the model. Most methodologies to date use molecular descriptors \cite{Keiser20MolReps} based on treating molecules as 2D objects -- graphs where the atoms are nodes and the bonds are edges. This leads to the extended connectivity fingerprint (ECFP) \cite{rogers2010extended} and more recent advances that extract the best possible representation of a 2D molecule graph using an end-to-end differentiable framework \cite{duvenaud2015convolutional}. 

Nonetheless, the physical mechanism that underlies biological activity is favourable interactions between local regions on the 3D surface of a molecule (pharmacophores) and residues in the receptor binding site. As such, one would expect that the 3D shape of the molecule would be a more appropriate input. Approaches that attempt to capture this such as Comparative Molecular Field Analysis \cite{cramer1988comparative,clark1990comparative} and Rapid Overlay of Chemical Structures \cite{rush2005shape} have been developed in the literature. However, those methods either require manual alignment \cite{cramer1988comparative,clark1990comparative} --  introducing bias -- or consider the similarity between the shapes of entire molecules \cite{masek1993molecular,rush2005shape}, overlooking the fact that it is often specific regions of the molecule that drive binding or determine physicochemical properties. Overcoming this limitation, one can coarse-grain the molecule into sites of salient interactions \cite{jenkins20043d}, but this requires prior insights on what are the important molecule-receptor interactions. Focusing on locality, Axen et al.~\cite{axen2017simple} developed an approach inspired by extended connectivity fingerprints, where the local 3D environments around each atom are mapped into a fixed length vector via hashing. However, this method only takes into account radial distances and leaves out angular information which is intuitively vital for understanding steric and electrostatic interactions between chemical substituents.

Ideally, we would utilize descriptors that are able to leverage the entire shape of the molecule using 3D atomic coordinates for property prediction. A challenge for designing such descriptors is in capturing precise geometric details of the molecular shape while remaining invariant to rotations/translations of the molecule, since they are physically identical.

Within condensed matter physics, this problem is routinely tackled as a question on how best to represent local atomic environments and has in particular received significant attention in the field of interpolating potential energy surfaces. An established mathematical formalism is that of Smooth Overlap of Atomic Positions (SOAP) \cite{bartok2013representing}. The key idea is to first represent each atomic environment using a sum of Gaussian densities, then ensure rotational invariance by integrating over all rotations (analytically tractable using the mathematics of spherical harmonics), and finally compute molecular similarity between two molecules by the similarity between the atomic environments that are the most similar. SOAP has found success in cracking challenging problems in materials science such as the phase behavior and defect structure of carbon \cite{caro2018growth}, boron \cite{deringer2018data} and silicon \cite{bartok2018machine}. Although SOAP has become the workhorse in computational physics, it has not been extensively tested and deployed in cheminformatics. Classifying docking decoys using SOAP \cite{bartok2017machine} has been reported, but the dataset involved suffers from acute bias due to the artificial enforcing of topological dissimilarity\cite{Rarey19}. Nonetheless, those results hint that SOAP could be a useful tool for QSAR modelling. This work seeks to more systematically investigate this by exploring the empirical utility of SOAP as a general purpose 3D QSAR method for the challenging prediction of experimental bioactivity.

In this paper, we will first describe an ML model utilising SOAP descriptors (SOAP-GP) and show that it can comfortably compete and outperform traditional fingerprint-based approaches as well as state-of-the-art graph neural networks on predicting binding affinity, even in challenging scaffold splits which address dataset bias \cite{wallach2018most,sieg2019need}. To further illustrate the usefulness of SOAP as an orthogonal descriptor, we include SOAP-GP in model ensembles using different representations and show an statistically significant improvement in performance.

\begin{figure}[!h] 
\centering
\includegraphics[width=\textwidth]{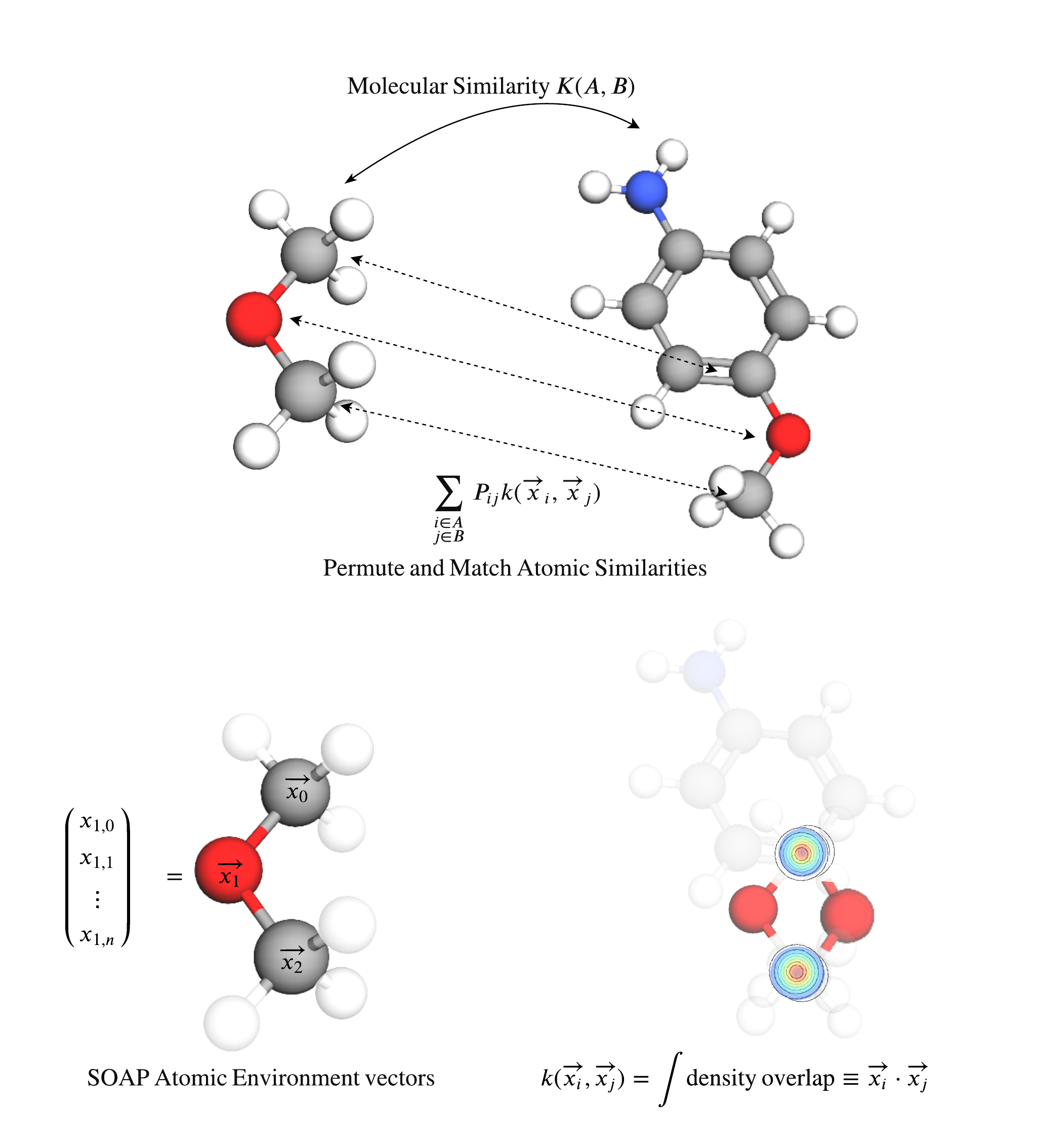}
\caption{\label{fig:soap} An illustration of how molecular similarity is defined by permuting and maximising the similarity of atomic environments between molecules.}
\end{figure}

\section{Results}\label{sec:results}
\subsection{SOAP-GP Model Description}\label{subsec:soapgp}

In the SOAP framework \cite{bartok2013representing}, the local atomic environment of an atom $\textbf{x}$ is represented by the sum of element-specific Gaussian densities centered on the positions of neighbourhood atoms. The ``similarity'' between two atomic environments is given by $k(\textbf{x}_{i}, \textbf{x}_{j}) = \textbf{x}_{i} \cdot \textbf{x}_{j}$, where the similarity function $k$ between environments represents the overlap of these neighbourhood densities, accounting for all possible element pairings, integrated over all coordinate system rotations (normalized so that self-similarity is unity). The procedure of integrating over all rotations ensures that any rotational transformation of the atomic coordinate system has no effect on $x_{i}$ or the similarity function. Using spherical harmonics and radial basis functions, this integral can be analytically computed as a truncated sum of coefficients -- the vector of these coefficients forms the descriptor $\textbf{x}$. This construction is invariant to rotations, translations, and permutations, and thus alignment-free. 

To find the geometric similarity between two molecules $A$ and $B$ (Fig \ref{fig:soap}), the local similarities of the best possible pairing of the atomic environments in $A$ and $B$ are used:
\begin{equation}\label{eq:rematch}
    K(A,B) = \displaystyle\sum_{\substack{i \in A\\ j \in B}}P_{ij}k(\textbf{x}_{i}, \textbf{x}_{j})
\end{equation}
where $P_{ij}$ is the $(i,j)$th element of the normalized permutation matrix $P$ that maximises $K$. This can be expressed as an optimal assignment problem and computed efficiently using a regularized entropy-matching approach, and is known as the ``REMatch'' similarity kernel~\cite{de2016comparing}. The `distance' between $A$ and $B$, which can be understood to be a measure of the geometric difference between these two molecules, can then be easily evaluated as

\begin{equation}\label{eq:distance}
    d(A, B) = \sqrt{2 - 2K(A, B)}
\end{equation}

The SOAP framework rigorously characterizes three-dimensional atomic environments and thus allows us to represent differences in molecular shape between individual molecules in a principled, alignment-free manner as a singular metric. SOAP has had great success as an atomic descriptor for machine learning interatomic potentials~\cite{bartok2018machine}, as well as for directly modelling material properties~\cite{Nyshadham2019}. Since the binding of ligands to proteins strongly depends on the three-dimensional interactions between the ligand and the receptor binding site, there is reason to expect that such a precise measure of molecular shape could also find success as an informative descriptor for predicting bioactivity.

This method differs in approach from conventional QSAR methods in that no explicit chemical descriptors (eg bonding, hybridization, aromaticity) are used at all in the featurisation of a molecule. Instead, chemical information is implicitly learned from the conformational shape of the molecule, from the coordinates of the atoms relative to one another, and completely encoded in the form of a numerical distance metric.

Such an approach naturally lends itself into the framework of kernel-based machine learning methods. By using nonlinear kernel functions to define distances between datapoints, these kernel methods implicitly project data into a higher-dimensional feature space where correlations could be more easily spotted. Well-known example of such methods are Support Vector Machines (SVMs) and more advanced Gaussian Process~\cite{rasmussen2005gp} (GP) models. Further discussion of kernel methods for QSAR can be found in this review by Muratov et\.al \cite{Muratov2020QSAR}. SOAP-based kernel models are regularly used for interpolating potential energy surfaces, and in this work we choose to implement a GP regression model.

GP Regression is a Bayesian ML method which searches over a probability distribution over functions which could model the data. The kernel $K(X_{i},X_{j})$ between data points is used as the covariance of the prior distribution over functions, and the training data is used to construct a likelihood. With Bayes' theorem this defines a posterior distribution for prediction. The model is trained by optimizing the kernel parameters in order to maximise the marginal likelihood of the distribution of functions which model the data.

\begin{figure}[!h]
\centering
\includegraphics[width=\textwidth]{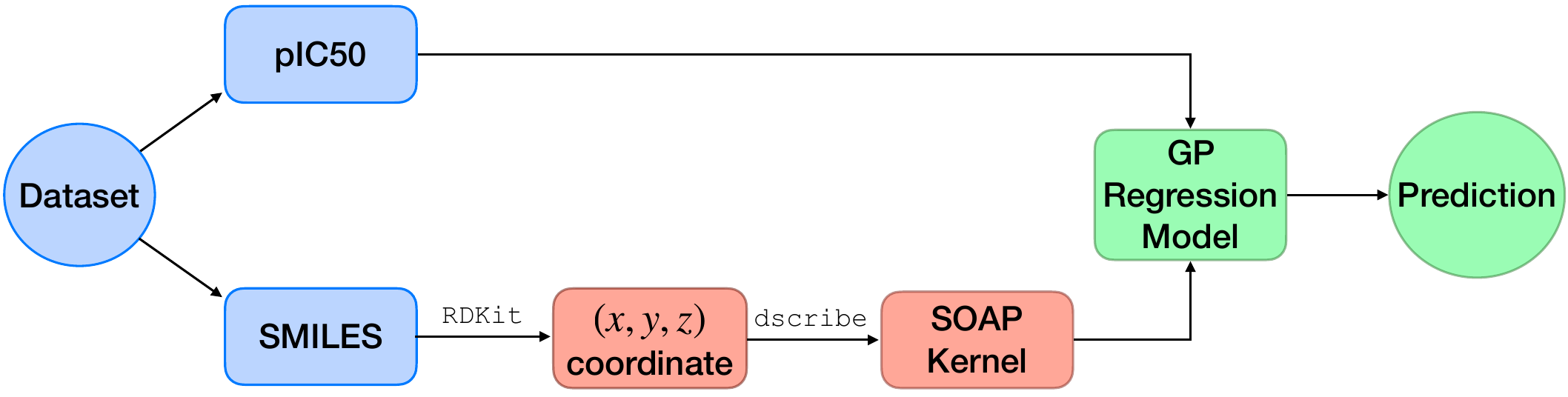}
\caption{\label{fig:soapgp} An overview of the SOAP-GP model implementation.}
\end{figure}

To incorporate smoothness and differentiability into the GP kernel and this way assisting in the learning of the model, we augment the REMatch distance $d(A,B)$ (Eq.~\ref{eq:distance}) with the $\nu=\frac{3}{2}$ Mat$\acute{\textrm{e}}$rn kernel $K^{\dagger}(A,B)$:

\begin{equation}
    K^{\dagger}(A,B) = \sigma^{2}\Big( 1+\frac{\sqrt{3}d}{\rho}\Big)\exp{\Big(-\frac{\sqrt{3}d}{\rho}\Big)}
\end{equation}
where $\sigma$ and $\rho$ are the kernel parameters (both initialised at unity) which are optimized to fit to training data. We refer to GP regression models utilizing SOAP features as SOAP-GP (Fig \ref{fig:soapgp}). 

\subsection{Comparative models}
The performance of our model was compared directly with that of several others which use representations of differing dimensionality and complexity. The intention of this exercise is not to establish an authoritative benchmark of QSAR model architectures, but as an empirical exploration of how SOAP-GP compares to representative example models which utilise particular molecular featurisations. Indeed, SOAP itself is merely one example of the many ways in which those in the field of materials science seek to precisely describe the atomic environments of molecular and crystal structures (eg atom-centered symmetry functions \cite{Behler2011}, FCHL \cite{Faber2018}, many-body tensor representations \cite{huo2017unified}). Recent work has shown that many of these representations are closely related as methods for representing atomic environments via symmetry-invariant atomic densities \cite{Willat19}, and we chose SOAP as a representative example of these approaches.

The industry standard approach for representing molecules is to use the extended connectivity fingerprint (ECFP), which considers molecules as 2D graphs and encodes the topological structural features of a molecule into a fixed-length binary bit string. ECFPs are a popular similarity search tool in drug discovery as the distance between two molecules can be simply defined as the Tanimoto distance between the bit strings \cite{Todeschini2012tanimoto}. We implement ECFPs in a random forest model (ECFP-RF), which is an established benchmark model for QSAR tasks.

An extension of ECFPs is to consider molecules as 3D structures instead of molecular graphs, which leads to the extended three-dimensional fingerprint (E3FP) \cite{Axen17e3fp}. The logic behind such an approach is that the 3D fingerprints are better able to encode stereochemistry and include information on the relationship between atoms close in space but distant in bond connectivity. The E3FP approach only considers radial distances between atoms in its featurisation, while SOAP features also encode angular information. Just like ECFPs, the similarity between two molecules can be calculated using the Tanimoto distance between their E3FP fingerprints.

For E3FPs there is no conventional model implementation - both random forest and Gaussian Process models were attempted and the GP models on average performed better so we from hereon utilise E3FPs in a GP framework (E3FP-GP) with the Mat{\'e}rn kernel in an identical fashion as the SOAP-GP except that the Tanimoto distances between molecular fingerprints are used in place of the SOAP REMatch distance. The difference in performance can be isolated to the quality of the molecular distance measures -- this would illustrate the importance of including angular information in featurising molecular shape.

Last but not least, we also consider the Directed Message Passing Neural Network (DMPNN) model~\cite{yang2019chemprop}, a state-of-the-art graph neural network which uses 2D molecular graphs explicitly encoding atomic and bond properties such as formal charge and conjugation as input features, usually as one-hot vectors. In graph neural networks, atom and bond features are combined with those of their neighbours via message-passing and convolutional embedding to construct a learnt global descriptor of a molecule, which is then passed through a neural network for property prediction. Graph neural networks have been gaining popularity in the cheminformatics community for property prediction \cite{Gilmer17mpnn, Feinberg18potentialnet}, and most recently the DMPNN model was utilised in a successful landmark deep learning search for novel antibiotics \cite{stokes2020antibiotic}. 

\begin{table}[!h]
  \caption{pIC$_{50}$ RMSE Results - the lowest RMSE for each dataset are bolded.} %(include R2?)
  \label{tbl:rand}
  \begin{tabular}{@{\extracolsep{4pt}}lrrrrr} 
    \hline 
    & \multicolumn{4}{c}{Random Split}\\
    \cline{2-5}
    Dataset & \multicolumn{1}{c}{ECFP-RF} & \multicolumn{1}{c}{E3FP-GP} & \multicolumn{1}{c}{DMPNN} & \multicolumn{1}{c}{SOAP-GP}\\
    \hline
    \hline
    A2a & $ 0.839\pm0.030 $& $ \textbf{0.793}\pm\textbf{0.034} $ & $ 0.993 \pm 0.062 $ & $ 0.924\pm0.064 $\\
    ABL1 & $ 0.848\pm0.018 $ & $ 0.843\pm0.019 $ & $ 0.965 \pm 0.030 $ & $ \textbf{0.798}\pm\textbf{0.017}$\\
    AChE & $ 0.784\pm0.006 $ & $ 0.868\pm0.007 $ & $ 0.783 \pm 0.011 $ & $ \textbf{0.761}\pm\textbf{0.009} $ \\
    Aurora-A & $ \textbf{0.830}\pm\textbf{0.010} $ & $ 0.900\pm0.008 $ & $ 0.842 \pm 0.008 $ & $ 0.844\pm0.009 $ \\
    B-raf & $ \textbf{0.712}\pm\textbf{0.008} $ & $ 0.786\pm0.008 $ & $0.778 \pm 0.010 $ & $ 0.720\pm0.008 $ \\
    Cannabinoid & $ 0.747\pm0.015 $ & $ 0.800\pm0.011 $ & $0.845 \pm 0.019$ & $ \textbf{0.716}\pm\textbf{0.011} $\\
    Carbonic & $ \textbf{0.659}\pm\textbf{0.016} $ & $ 0.670\pm0.013 $ & $0.702 \pm 0.023$ & $ 0.839\pm0.095 $ \\
    Caspase & $ \textbf{0.587}\pm\textbf{0.008} $ & $ 0.662\pm0.009 $ & $0.597 \pm 0.012$ & $ 1.096\pm0.061 $ \\
    Coagulation & $ \textbf{0.909}\pm\textbf{0.010}$ & $ 1.010\pm0.009 $ & $1.019 \pm 0.022$ & $ 0.984\pm0.037 $\\
    COX-1 & $ 0.729\pm0.013 $ & $ 0.744\pm0.013 $ & $0.732 \pm 0.011$ & $ \textbf{0.706}\pm\textbf{0.013} $ \\
    COX-2 & $ 0.790\pm0.007 $ & $ 0.826\pm0.007 $ & $0.804 \pm 0.012$ & $ \textbf{0.762}\pm\textbf{0.007} $\\
    Dihydrofolate & $ \textbf{0.799}\pm\textbf{0.025} $ & $ 0.849\pm0.019 $ & $0.890 \pm 0.023$ & $ 0.811\pm0.021 $ \\
    Dopamine & $ \textbf{0.747}\pm\textbf{0.013} $ & $ 0.816\pm0.014 $ & $0.921 \pm 0.020$ & $ 0.777\pm0.017 $ \\
    Ephrin & $ 0.722\pm0.011 $ & $ 0.749\pm0.007 $ & $0.719 \pm 0.009$ & $ \textbf{0.701}\pm\textbf{0.008} $\\
    erbB1 & $ \textbf{0.756}\pm\textbf{0.003} $ & $ 0.818\pm0.005 $ & $0.748 \pm 0.010$ & $ 0.772\pm0.003 $ \\
    Estrogen & $ 0.691\pm0.005 $ & $ 0.697\pm0.005 $ & $0.670 \pm 0.007$ & $ \textbf{0.633}\pm\textbf{0.007} $\\
    Glucocorticoid & $ \textbf{0.612}\pm\textbf{0.010} $ & $ 0.663\pm0.008 $ & $0.691 \pm 0.008$ & $ 0.613\pm0.007 $ &\\
    Glycogen & $ \textbf{0.743}\pm\textbf{0.007} $ & $ 0.788\pm0.008 $ & $0.806 \pm 0.009$ & $ 0.769\pm0.006 $ \\
    HERG & $ 0.610\pm0.006 $ & $ 0.679\pm0.005 $ & $0.615 \pm 0.007$ & $ \textbf{0.569}\pm\textbf{0.005} $\\
    JAK2 & $ \textbf{0.672}\pm\textbf{0.007} $ & $ 0.737\pm0.007 $ & $0.719 \pm 0.007$ & $ 0.683\pm0.009 $ \\
    LCK & $ 0.829\pm0.010 $ & $ 0.867\pm0.012 $ & $0.918 \pm 0.021$ & $ \textbf{0.827}\pm\textbf{0.010} $ \\
    Monoamine & $ \textbf{0.676}\pm\textbf{0.007} $ & $ 0.680\pm0.008 $ & $0.724 \pm 0.012$ & $ 0.680\pm0.009 $ \\
    opioid & $ 0.729\pm0.011 $ & $ 0.781\pm0.018 $ & $0.748 \pm 0.020$ & $ \textbf{0.692}\pm\textbf{0.015} $\\
    Vanilloid & $ 0.724\pm0.006 $ & $ 0.774\pm0.006 $ & $0.744 \pm 0.008$ & $ \textbf{0.720}\pm\textbf{0.006} $ \\
    \hline
    \end{tabular}
\end{table}
\begin{table} [!h]
\begin{tabular}{@{\extracolsep{4pt}}lrrrrr}
\hline
  %\caption{RMSE Results - E3FP 1024bits, ECFP 2048 bits}
    & \multicolumn{4}{c}{Scaffold Split}\\
    \cline{2-5}
    Dataset & \multicolumn{1}{c}{ECFP-RF} & \multicolumn{1}{c}{E3FP-GP} & \multicolumn{1}{c}{DMPNN} & \multicolumn{1}{c}{SOAP-GP}\\
    \hline
    \hline
    A2a & $ 1.113\pm0.087 $ & $ 1.134\pm0.128 $ & $1.434 \pm 0.194$ & $ \textbf{1.028}\pm\textbf{0.065} $ \\
    ABL1 & $ \textbf{0.933}\pm\textbf{0.036} $ & $ 0.971\pm0.047 $ & $1.069 \pm 0.043$ & $ 0.951\pm0.050 $ \\
    AChE & $ 0.990\pm0.023 $ & $ 1.045\pm0.025 $ & $0.994 \pm 0.030$ & $ \textbf{0.952}\pm\textbf{0.022} $ \\
    Aurora-A & $ \textbf{0.928}\pm\textbf{0.025} $ & $ 1.011\pm0.021 $ & $0.953 \pm 0.029$ & $ 0.942\pm0.017 $ \\
    B-raf & $ 0.866\pm0.038 $ & $ 0.916\pm0.038 $ & $0.959 \pm 0.032$ & $ \textbf{0.841}\pm\textbf{0.035} $ \\
    Cannabinoid & $ 0.874\pm0.026 $ & $ 0.943\pm0.028 $ & $0.967 \pm 0.027$ & $ \textbf{0.827}\pm\textbf{0.022} $ \\
    Carbonic & $ \textbf{0.682}\pm\textbf{0.032} $ & $ 0.816\pm0.044 $ & $0.809 \pm 0.060$ & $ 0.689\pm0.049 $ \\
    Caspase & $ \textbf{0.721}\pm\textbf{0.040} $ & $ 0.764\pm0.027 $ & $0.770 \pm 0.025$ & $ 0.922\pm0.063 $ \\
    Coagulation & $ 0.996\pm0.014 $ & $ 1.076\pm0.023 $ & $1.100 \pm 0.025$ & $ \textbf{0.989}\pm\textbf{0.025} $ \\
    COX-1 & $ 0.793\pm0.017 $ & $ 0.789\pm0.014 $ & $\textbf{0.768} \pm \textbf{0.008}$ & $ 0.781\pm0.009 $ \\
    COX-2 & $ 1.008\pm0.039 $ & $ 1.009\pm0.031 $ & $1.010 \pm 0.037$ & $ \textbf{0.960}\pm\textbf{0.033} $ \\
    Dihydrofolate & $\textbf{0.914}\pm\textbf{0.058} $ & $ 0.938\pm0.051 $ & $1.012 \pm 0.044$ & $ 0.967\pm0.057 $ \\
    Dopamine & $ \textbf{0.869}\pm\textbf{0.020} $ & $ 0.882\pm0.018 $ & $0.940 \pm 0.020$ & $ 0.894\pm0.020 $ \\
    Ephrin & $ \textbf{0.881}\pm\textbf{0.018} $ & $ 0.908\pm0.028 $ & $0.904 \pm 0.025$ & $ 0.882\pm0.021 $ \\
    erbB1 & $ 0.888\pm0.013 $ & $ 0.947\pm0.012 $ & $\textbf{0.864} \pm \textbf{0.010}$ & $ 0.891\pm0.007 $\\
    Estrogen & $ 0.795\pm0.018 $ & $ 0.786\pm0.015 $ & $0.744 \pm 0.014$ & $ \textbf{0.708}\pm\textbf{0.011} $ \\
    Glucocorticoid & $ 0.742\pm0.023 $ & $ 0.790\pm0.024 $ & $0.859 \pm 0.022$ & $ \textbf{0.738}\pm\textbf{0.014} $ \\
    Glycogen & $ \textbf{0.869}\pm\textbf{0.021}$ & $ 0.910\pm0.022 $ & $0.963 \pm 0.022$ & $ 0.906\pm0.020 $ \\
    HERG & $ 0.690\pm0.018 $ & $ 0.747\pm0.021 $ & $0.706 \pm 0.023$ & $ \textbf{0.656}\pm\textbf{0.018} $ \\
    JAK2 & $ 0.746\pm0.010 $ & $ 0.803\pm0.013 $ & $0.783 \pm 0.019$ & $ \textbf{0.738}\pm\textbf{0.021} $ \\
    LCK & $ \textbf{0.909}\pm\textbf{0.014} $ & $ 0.954\pm0.018 $ & $1.056 \pm 0.030$ & $ 0.918\pm0.012 $ \\
    Monoamine & $ 0.818\pm0.022 $ & $ \textbf{0.813}\pm\textbf{0.023} $ & $0.927 \pm 0.030$ & $ 0.817\pm0.023 $\\
    opioid & $ 0.781\pm0.032 $ & $ 0.797\pm0.031 $ & $0.811 \pm 0.028$ & $ \textbf{0.747}\pm\textbf{0.021} $ \\
    Vanilloid & $ 0.770\pm0.018 $ & $ 0.814\pm0.018 $ & $0.826 \pm 0.026$ & $ \textbf{0.762}\pm\textbf{0.015} $ \\
    \hline
    \end{tabular}
\end{table}

\subsection{RMSE Performance}\label{subsec:performance}
To find out the performance of these models we used IC$_{50}$ datasets for 24 diverse protein targets extracted from ChEMBL which have been previously investigated in several screening and modelling studies \cite{Bender2018, Bender2019}. IC$_{50}$ measures the concentration of a compound required for the inhibition of a target to drop by 50\% - the IC$_{50}$ (or pIC$_{50}$ = $-\log_{10}$IC$_{50}$) values are a direct metric of ligand-protein binding affinity, and modelling these values is thus an appropriate challenge for comparing QSAR models. The datasets are further filtered to remove large compounds beyond the scope of small molecule drug discovery.

The above models are compared by evaluating the root-mean-square errors (RMSE) of their predictions on the same train/test splits of the datasets. Besides random splitting, we also evaluate on these datasets using scaffold split, which ensures that training and test sets do not share molecules with similar Bemis-Murcko scaffolds. This method of splitting better simulates the real-life drug discovery cycle where prior activity data only exists for a class of chemical compounds that are different from those that are being evaluated, in other words posing a greater extrapolation challenge. All results are from the mean and standard errors from 15 independent runs.

With random splitting (Table~\ref{tbl:rand}), the well-established ECFP-RF method demonstrates its effectiveness, outperforming the others on 12 of the 24 tasks with SOAP-GP coming in second at 11 out of 24, leaving only the A2a subset for E3FP-GP. A similar picture is seen under scaffold splitting where in this case SOAP-GP does best on 12 of the 24 tasks, with 9 for ECFP-RF, only two for DMPNN, and one for E3FP-GP. A more challenging test scenario, the scaffold split results in overall higher RMSEs and standard deviations. In all cases the model predictions are far above the typical recorded error of $\pm0.5$ log units~\cite{Kallioski13}, illustrating the general difficulty in modelling pIC$_{50}$ values.

These results show that SOAP-GP, utilising out-of-the-box open-source descriptors of three-dimensional molecular shape from condensed matter physics, is competitive with both conventional and current state-of-the-art ML QSAR models. In particular, comparing SOAP-GP against E3FP-GP suggests that merely accounting for radial distances is an insufficiently informative description of shape. The informational richness of the SOAP descriptor in containing extensive angular information about atomic environments, required for its original purpose of fitting interatomic potentials, allows SOAP-GP to far better model binding affinity.

\subsection{Ensembling Representations}
Despite the showcased competitive capabilities of SOAP-GP, we do not propose that SOAP-GP should become a new paradigm in cheminformatics QSAR, nor indeed that any sole representation/model should be. From this dataset of 24 targets alone it can already be observed that model performance can vary substantially and that it is hard to know a priori which model would do best. %We argue that this is the case generally in bioactivity prediction and it is unhelpful to solely rely on any particular `state of the art' model or molecular featurisation for QSAR modelling.

%Our argument is specifically for the learning of accurate ligand-binding affinity. When conducting a vast virtual screen of a molecular library or attempting to classify activity from a large high-throughput-screen, practical considerations such as time constraints and the memory scaling behaviour of models should dictate the choice of model. However, high-accuracy IC$_{50}$-accuracy activity data rarely exceeds $N=10^{4}$ in size \cite{Cherkasov2014QSAR}, and in this regime computational constraints are typically minor compared to demands on predictive performance.

In this scenario, a straightforward way to achieve improved performance is to combine QSAR models in an ensemble learning approach where the predictions from several models are averaged to give better results \cite{Sagi2018ensemble}. Such an approach is only successful if there is sufficient diversity such that each model captures trends in the dataset that are neglected by the others. The power of model ensembling lies not merely via the principle of `strength in numbers', but `strength in diversity'.

While model ensembling in QSAR has been explored before, it is often done in the context of ensembling different model architectures on the same representation. Ensembling diverse representations, however, is less common. Unlike the conventional applications of machine learning, chemistry lends itself to rich and diverse featurisations and this fact should be taken advantage of. Models trained on hybridization states and stereochemistry will capture distinct effects from those trained using conformational shapes, and we suggest that the 3D atomic environments described by SOAP allow it to serve as a useful descriptor orthogonal to those commonly used. %and these differences should be more significant than the particular architecture that is employed.

\begin{figure}[!h]

\centering
\includegraphics[width=\textwidth]{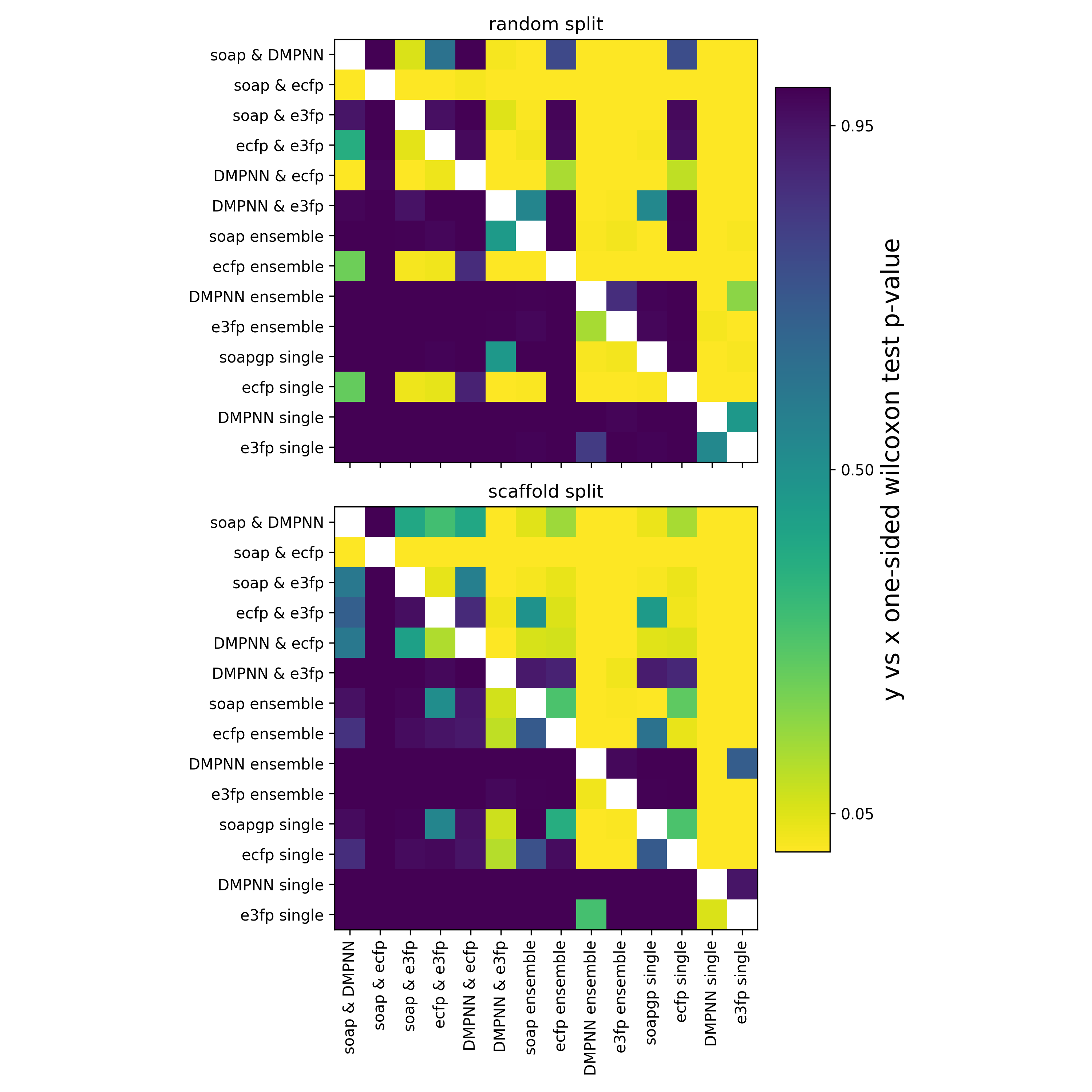}
\caption{\label{fig:ensemble_mat} Ensembling diverse representations is superior to ensembling similar representations regardless of model architecture. Colour indicates the $p$-value for the one-sided Wilcoxon signed rank test with alternate hypothesis ``model $y$ has a lower mean RMSE than model $x$''. Small $p$-values (yellow) indicate that the null hypothesis ``model $y$ does \textit{not} have a lower mean RMSE than model $x$'' can be rejected.}
\end{figure}

We demonstrate this by comparing the performance of ensembles pairing models of diverse representations, as well as only single non-ensembled models, using the Wilcoxon signed-rank test. The Wilcoxon signed-rank test is a non-parametric paired difference test used to compare samples from two distributions, statistically testing whether or not the difference between the two distributions are centered around zero -- this test has been previously used to evaluate model performance on bioactivity prediction~\cite{Mayr2018comparison}. We treat each model's RMSEs on the 24 IC$_{50}$ datasets as a single statistical sample, and perform a one-sided test between $(x,y)$ pairs of model RSMEs with the null hypothesis "model $y$ has a higher or equal mean RMSE to model $x$" versus the alternative hypothesis “model $y$ has lower mean RMSE than model $x$". The $p$-values for the tests are evaluated, and plotted as a matrix in Fig~\ref{fig:ensemble_mat}. Bright yellow patches indicate that the null hypothesis has a small $p$-value and can be rejected, statistically confirming that model $y$ (listed on the vertical axis) indeed has a lower mean RMSE than model $x$.

It can be seen that ensembling diverse representations almost always statistically outperforms ensembling the same representation, which in turn tends to be better than the single models on their own. These differences are most accentuated in the more realistic scaffold split scenario. The ensemble of SOAP-GP and ECFP-RF is statistically better performing than any of the other possible combinations. This is not entirely surprising given that these were the two best-performing single models on their own, but it demonstrates that the trends learnt by the two models complement one another, that combining 2D topological information with precise 3D atomic features can push the frontier of QSAR modelling. Additionally, a reinforcement of our previous observation in comparing SOAP-GP to ECFP-RF can also be seen -- the two single models cannot be statistically distinguished in a scaffold split scenario, and only for random splits can we meaningfully say that ECFP-RF is the best performing single model.

\section{Discussion}
Before concluding, we would like to discuss several limitations of our approach. It is a great surprise that SOAP-GP was able to perform as well as it does even though only a single conformer is used as the three-dimensional molecular shape for the generation of the SOAP descriptors. In reality molecules exist in equilibria between multiple conformers, Boltzmann distributed by differing free energies due to electrostatic, steric, and orbital interactions. How the model performance varies with conformer generation methodology, as well as whether or not it could be improved by including multiple conformers, is the subject of further investigation.

In addition, while it is evident that the incorporation 3D atomic environments in SOAP-GP allows it to correlate molecular shape to binding affinity, it is not easy to understand what kind of three-dimensional shape features the model uses to make its predictions. Not only do the conformational shapes of the input data need to be assessed and compared, but also the three-dimensional shape and interactions at the binding site of the protein target need to be considered. This requires precise investigation and should be the subject of future work.

The competitive performance of SOAP-GP implies that the SOAP distance $d$ (Eq.~\ref{eq:distance}), after fitting via the GP kernel, can also serve as an application-specific, property-sensitive measure of the `distance' between molecules. While the use of SOAP for the embedding and visualisation of the abstract space spanned by atomic structures has been investigated in a materials science context~\cite{reinhardt2019ti02,bartok2017machine}, this has not yet been done specifically in the domain of medicinal chemistry on drug-like molecules. 

The success of SOAP-GP in modelling ligand-protein binding affinity suggests that many other atomic/structural descriptors from the field of machine learning force fields (such as FCHL~\cite{Faber2018}, many-body tensor representations~\cite{huo2017unified}), as well as the kaleidoscopic model architectures (such as SchNet~\cite{schnet}, ANI-1~\cite{2017ANI}) that utilise those descriptors for the purpose of predicting quantum energies, have the potential to also be useful for QSAR modelling. We foresee a great deal of fruitful cross-fertilization between the cheminformatics community and that of interpolating potential energy surfaces in the future.

\section{Conclusion}
We described SOAP-GP, an alignment-free 3D QSAR method which employs a GP model on the intermolecular similarity between local atomic environments featurized using open-source SOAP descriptors borrowed from condensed matter physics. The performance of this model was empirically compared with a 2D fingerprint-based random forest model, a 3D fingerprint-based GP, as well as a state-of-the-art graph neural network, on 24 pIC$_{50}$ regression tasks from ChEMBL. We showed that SOAP-GP, utilizing out-of-the-box open-source descriptors, is competitive with all of these on both random and challenging scaffold splits.

We further demonstrate the utility of SOAP descriptors by creating ensembles of models paired with one another and comparing their performance using the Wilcoxon signed-rank test. We find that ensembles with diverse representations statistically outperform those with the same representation, and that SOAP-GP combined with ECFP-RF has the strongest performance, showcasing the value of combining 2D features with 3D atomic environment descriptors in capturing information relevant to predicting binding affinity.

These results show that capturing 3D atomic environments from conformers, where there has been much prior work from the condensed matter community, has value for QSAR modelling as an orthogonal descriptor to traditional approaches. We anticipate that methods from the field of interpolating potential energy surfaces will continue to be a source of inspiration to the cheminformatics community and look forward to further cross-disciplinary transfer of ideas.
%Although much research is done on continually developing novel and improved QSAR models on a competitive basis, by instead taking a collaborative approach and complementing models with one another to exploit feature-rich chemistry, enhanced QSAR for medicinal chemistry can be achieved.

\section{Experimental Details}\label{sec:experiment}
\subsection{Datasets details}\label{subsec:datasets}
The IC$_{50}$ datasets used in this work were extracted from ChEMBL database version 23 and had previously undergone filtering to only include precise measurements. However, we additionally found that in many cases they also contained large compounds such as glycans and oligopeptides which are unreasonable candidates for a small molecule drug discovery campaign. We filter the dataset to only keep molecules with molecular weight below 500 daltons (as per Lipinski's rules) which reduces the datasets by 19\% on average in size (Table~\ref{table:bender_datasets}).

\begin{table}[!h]
\caption{ChEMBL bioactivity data used in this study}
\centering
\label{table:bender_datasets}
\begin{tabular*}{\textwidth}{ll@{\extracolsep{\fill}}cc}
\hline
ChEMBL target preferred name & Abbreviation & Initial Size & Size after filtering\\ 
\hline
\hline
Alpha-2a adrenergic receptor&A2a & 203 & 166\\
tyrosine-protein kinase ABL &ABL1 & 773 & 536\\
acetylcholinesterase &AChE & 3159 & 2491\\
serine/threonine-protein kinase aurora-A &Aurora-A & 2125 & 1612\\
serine/threonine-protein kinase B-raf &B-raf & 1730 & 824\\
cannabinoid CB1 receptor &Cannabinoid & 1116 & 820\\
carbonic anhydrase II &Carbonic & 603 & 556\\
caspase-3&Caspase & 1606 & 1362\\
thrombin &Coagulation & 1700 & 862\\
cyclooxygenase-1 &COX-1 & 1343 & 1278\\
cyclooxygenase-2 &COX-2 & 2855 & 2704\\
dihydrofolate reductase &Dihydrofolate & 584 & 548\\
dopamine D2 receptor &Dopamine & 479 & 405\\
norepinephrine transporter &Ephrin & 1740 & 1716\\
epidermal growth factor receptor erbB1 &erbB1 & 4868 & 3598\\
estrogen receptor alpha &Estrogen & 1705 & 1546\\
glucocorticoid receptor &Glucocorticoid & 1447 & 1077\\
glycogen synthase kinase-3 beta &Glycogen & 1757 & 1655\\
HERG&HERG & 5207 & 4042\\
tyrosine-protein kinase JAK2 &JAK2 & 2655 & 2252\\
tyrosine-protein kinase LCK &LCK & 1352 & 954\\
monoamine oxidase A&Monoamine & 1379 & 1344\\
Mu opioid receptor&opioid & 840 & 611\\
manilloid receptor&Vanilloid & 1923 & 1656\\
\hline
\end{tabular*}
\end{table}

For SOAP-GP, ECFP-RF, and E3FP-GP, the datasets are split 80/20 into train/test sets and for the DMPNN models the split is 70/10/20 for train/validation/test sets. The random split results are given as the mean results from 15 runs.

When evaluating datasets by scaffold split, molecules are binned by Murcko scaffold (evaluated using \texttt{RDKit}). Bins larger than half of the required test set size are placed in the training/validation set and all remaining bins are distributed randomly such that the required train/test split sizes are met. The scaffold split results are given as the mean results from 15 runs using different random seeds for the distribution of scaffolds.

\subsection{Model implementation}
Computationally, SMILES strings of molecules are converted into $(x,y,z)$ atomic coordinates using the ETKDG conformer generation method \cite{Riniker15ETKDG} implemented in \texttt{RDKit} \cite{rdkit}. Only one conformer is generated. SOAP descriptors and kernels were computed from the resultant atomic coordinates using the \texttt{soapxx} and \texttt{dscribe} packages \cite{soapxx,dscribe}, with the basis function parameters $ n_{max}=12, l_{max}=8$. Two sets of SOAP descriptors with $r_{cut}=3.0\textup{\AA}, \sigma=0.2\textup{\AA}$ and $r_{cut}=6.0\textup{\AA}, \sigma=0.4\textup{\AA}$ were evaluated and concatenated for each molecule. These hyperparameters were chosen based on standard values used for structure modelling with SOAP in condensed matter physics. For the REMatch kernel, the entropy regularization parameter $\alpha$ was manually set to $0.5$ based on predictive performance with a convergence threshold of $10^{-6}$. The GP model itself was implemented using \texttt{GPFlow}.

The relatively large value of $\alpha$ was chosen to make the resultant kernel intermediate between the average and best-match molecular kernels~\cite{de2016comparing}. This was motivated by the fact that the average kernel was shown to be an appropriate choice for modelling extensive properties (those that can be decomposed into atomic contributions), while the best-match kernel performs better for intensive properties~\cite{bartok2018machine}. Choosing an intermediate value of $\alpha$ is a compromise of the two approaches, and allows the model to better generalise to modelling different properties.   

ECFP fingerprints were generated with 1024 bits and a radius of 3 using \texttt{RDKit}, while E3FP fingerprints were generated also with 1024 bits using the \texttt{e3fp} package \cite{Axen17e3fp}.

The DMPNN model was implemented using the \texttt{chemprop} package. The training procedure regarding the molecular features used as well as the initial hyperparameter optimization was done following the guidelines from \cite{yang2019chemprop}.

\subsection{Code}
Code for generating the SOAP features and implementing the SOAP-GP model can be found in the GitHub repo \texttt{soapgp}  \cite{McCorkindale2020SOAPGP}.

\section{Acknowledgements}
We thank G\'{a}bor Cs\'{a}nyi for insightful discussion. WM acknowledges the support of the Gates Cambridge Trust. AAL acknowledges the Winton Programme for the Physics of Sustainability. Computations were performed at the CSD3 High Performance Computing Service at the University of Cambridge.

\bibliography{reference.bib} 

\end{document}